\documentclass[11pt]{article}
\usepackage{moriond,epsfig}

\def\Journal#1#2#3#4{{#1} {\bf #2}, #3 (#4)}

\def\NPB{{\em Nucl. Phys.} B}
\def\PLB{{\em Phys. Lett.}  B}
\def\PRL{\em Phys. Rev. Lett.}
\def\PRD{{\em Phys. Rev.} D}

\def\JHEP{\em JHEP}

\begin{document}
\vspace*{4cm}
\title{Measurements of Rare {\boldmath $B$} Decays at Tevatron}

\author{ Masato Aoki }

\address{Fermi National Accelerator Laboratory,\\
Batavia, Illinois 60510, USA \\
(On behalf of the CDF and D\O\ Collaborations)
}

\maketitle\abstracts{
  Both CDF and D\O\ experiments have been searching for evidence of physics beyond the standard model (SM) using the Tevatron $p\bar{p}$ collider at Fermilab.
  We report on recent searches in the $B$ flavor sector, especially decays via flavor changing neutral current processes (FCNC), $B^0_{(s)}\to e^+\mu^-$ and $B^0_{s}\to\mu^+\mu^-$, at the Tevatron.
}

\section{Introduction}

The branching fraction of a rare decay mode is an interesting quantity to measure
because the contribution from physics beyond the SM may be sizable in rare decay modes.
In order to be able to observe rare heavy flavor decays it is essential to produce a sufficient number of bottom hadrons.
The bottom anti-bottom production cross-section $\sigma(b\bar{b})$ at the
Tevatron is O(10$^{5}$) larger than production in $e^{+}e^{-}$ colliders at
the $\Upsilon(4s)$ or $Z^{0}$ energy scale.
The large production of all kinds of $b$-hadrons at the Tevatron offers the 
opportunity to study rare decays also in the $B^0_{s}$ and $b$-baryon sectors.
On the other hand the inelastic cross section is $10^3$ times higher than $\sigma(b\bar{b})$ requiring very selective and efficient triggers.
Therefore, interesting events must be extracted from a high track multiplicity
environment, and detectors need to have very good tracking, vertex resolution, wide acceptance, good particle identification
and highly selective triggers.

In this article we report on searches for rare $B$ decays via FCNC process, $B^0_{(s)}\to e^+\mu^-$ and $B^0_{s}\to\mu^+\mu^-$, using the Tevatron $p\bar{p}$ collider at Fermilab.

\section{{\boldmath $B^0_{(s)} \to e^+ \mu^-$} Decays}
In the SM, lepton number and lepton flavor are conserved; therefore the decays such as $B^0_{(s)}\rightarrow e^+ \mu^-$ are forbidden.
However the observation of neutrino oscillations indicates that
lepton flavor is not conserved while lepton flavor violating (LFV) decays in the charged sector have not been observed yet.
The grand-unification theory by J. Pati and A. Salam predicts a new
interaction to mediate transitions between leptons and quarks via exchange of
spin-1 gauge bosons, which are called Pati-Salam leptoquarks (LQ), that carry both
 color and lepton quantum numbers~\cite{pati-salam}.
The lepton and quark components of the leptoquarks are not
necessarily from the same generation~\cite{scott,Blanke}, and
the decays  $B^0_{s}\rightarrow e^+ \mu^-$ and $B^0\rightarrow e^+ \mu^-$
can be mediated by different types of leptoquarks.

\begin{figure}[htbp]
\center{
\psfig{figure=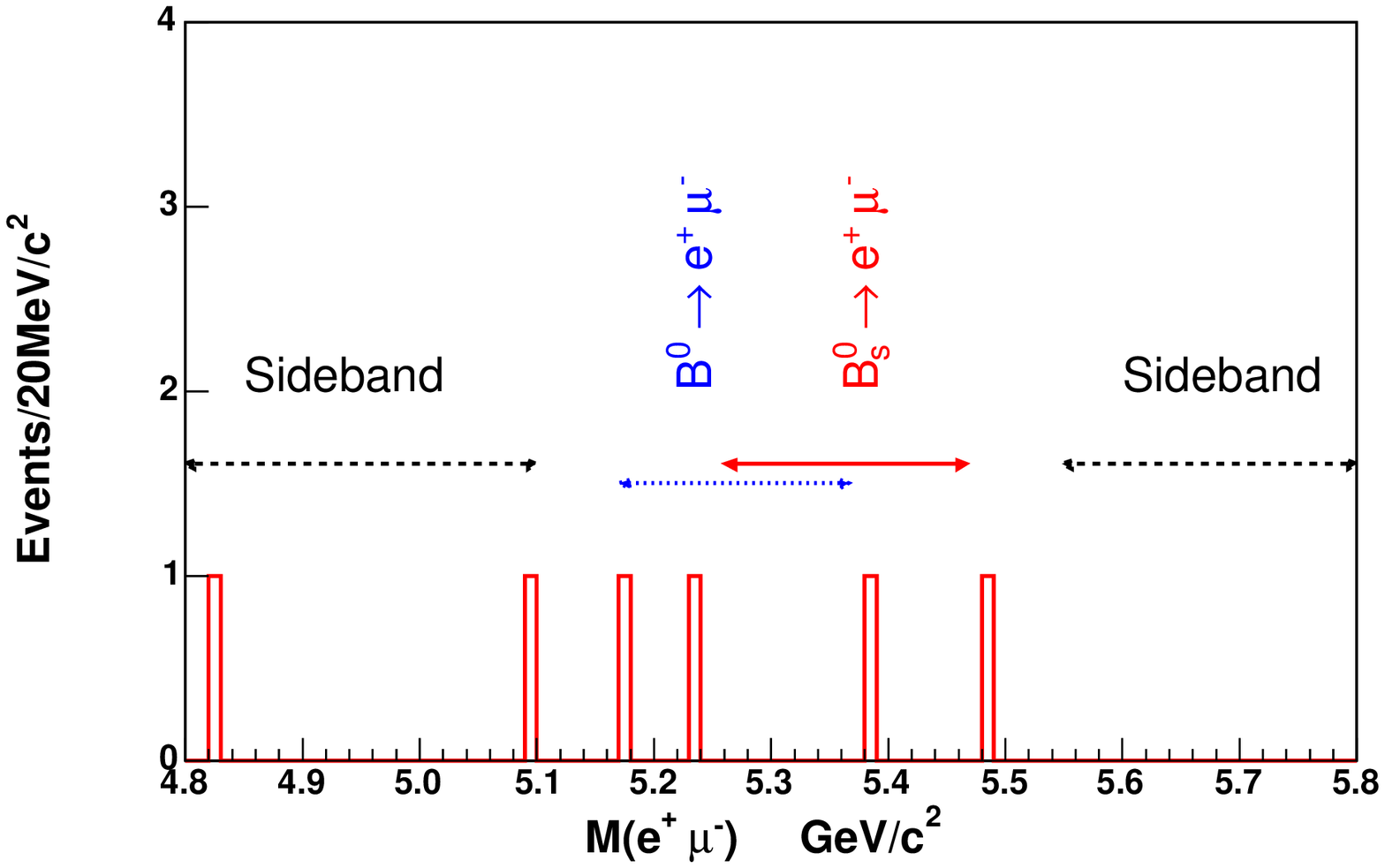,width=0.45\textwidth}
\psfig{figure=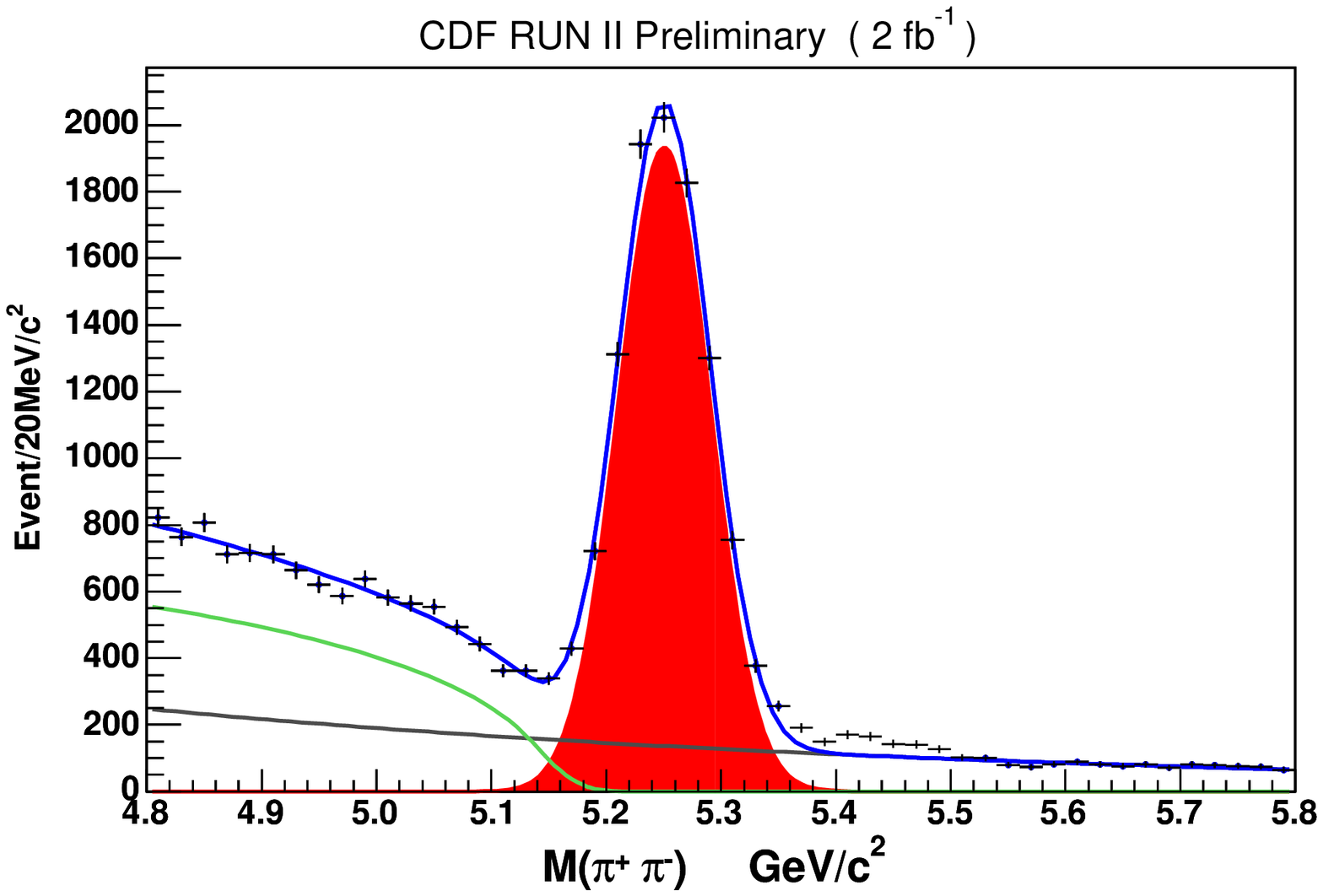,width=0.45\textwidth}
\caption{$e^+ \mu^-$ invariant mass distribution(left) and the reference channel $B^0 \to K^+\pi^-$ (right) in data.
\label{fig:b2emu}}
}
\end{figure}

CDF reported on  searches for the LFV decays $B^0_{(s)}\rightarrow e^+\mu^-$ using $2~{\rm fb}^{-1}$ of Run~I\hspace{-0.1em}I data~\cite{cdfb2emu}.
Data sample used in the searches was taken with the two displaced-track trigger, in which two oppositely-charged tracks are required to have a transverse momentum $p_T>2~{\rm GeV}/c$ and an impact parameter~\cite{IP} $0.1 < d_0 < 1~mm$, and it is also required that the scalar sum of the transverse momenta of the two tracks is greater than $5.5~{\rm GeV}/c$,
the difference in the azimuthal angles of  the tracks is $20^\circ <\Delta \varphi<135^\circ$,
and a transverse decay length~\cite{lxy}  is $L_{xy}>200$ $\mu$m.
In the off-line analysis, additionally the $B^0_{(s)}$ isolation, the pointing angle and the transverse~decay~length were required to be consistent with those of $B^0_{(s)}\rightarrow e^+\mu^-$ decays.
These thresholds were optimized in an unbiased way to obtain the best sensitivity for the searches~\cite{punzi}.
To identify electrons, both the specific ionization ($\mathit{dE/dx}$)~\cite{eID}
measured in the central~drift~chamber and the transverse and longitudinal shower shapes as measured in the central~electromagnetic~calorimeter were used.
The electron identification efficiency is $\sim70\%$ while the muon identification is fully efficient with $p_T>2~{\rm GeV}/c$ in the central muon detector.
Search windows in $e^+ \mu^-$ invariant mass were defined to be (5.262--5.477)~GeV/c$^2$ for  $B^0_{s} \rightarrow e^+ \mu^-$ and (5.171--5.387)~GeV/c$^2$  for $B^0 \rightarrow e^+ \mu^-$.
These correspond to a window  around the nominal values of the
$B^0_s$  and  $B^0$ masses~\cite{PDG} of approximately $\pm 3 \sigma_{m}$.
The background contributions considered include combinations of
random track pairs and partial $B$ decays that accidentally meet the
selection requirement 
and hadronic two-body $B$ decays in which both final
particles are misidentified as leptons.
The random track contribution was evaluated by extrapolating the normalized
number of events found in the sidebands to the signal region.
The double-lepton misidentification rate was determined by applying electron and muon
misidentification probabilities to the number of two-body decays found in the
search window.
Figure~\ref{fig:b2emu} shows
the invariant mass distribution  for $e^+ \mu^-$ candidates.
One event in the  $B^0_s$ mass window, and two events in the $B^0$ mass window were observed.
These numbers are consistent
with the estimated total background of  $0.8\pm 0.6$ events in  the $B^0_s$  search window, and $0.9 \pm 0.6$ in the $B^0$  window.
$B^0 \to K^+ \pi^-$ decays were used as a reference channel to set a limit on $\mathcal{B}(B^0_{(s)}\to e^+\mu^-)$.
Using the same selection criteria except lepton identification, $6387 \pm  214$ events of $B^0 \to K^+ \pi^-$ decays were observed as shown in Fig.~\ref{fig:b2emu}.
Resulting upper limits were $\mathcal{B}(B^0_s \to e^+ \mu^-) < 2.0~(2.6) \times 10^{-7}$ and $\mathcal{B}(B^0 \to e^+ \mu^-) < 6.4~(7.9) \times 10^{-8}$ at 90\% (95\%) confidence level (C.L.). 
Using the limits on the branching fractions, the masses of the corresponding
Pati-Salam leptoquarks were calculated to be ${\rm M_{LQ}}(B^0_s \to e^+ \mu^-) >  47.8~(44.9)\; {\rm TeV/c^2}$ and ${\rm M_{LQ}}(B^0 \to e^+ \mu^-) >  59.3~(56.3)\; {\rm TeV/c^2}$ at 90 (95)\% C.L.
These are the best limits in the world to date.

\vspace{-0.1cm}

\section{{\boldmath $B^0_s \to \mu^+ \mu^-$} Decays}

Branching fraction of the pure leptonic FCNC processes like
$B^0_{s} \rightarrow \mu^{+} \mu^{-}$ in the SM is heavily suppressed~\cite{Bla}:
$\mathcal{B}(B^0_{s} \rightarrow \mu^{+} \mu^{-})=(3.35\pm 0.32) \times 10^{-9}$.
$B^0 \to \mu^+\mu^-$ decay is further suppressed by a factor $| V_{td}/V_{ts} | ^{2}$ in the CKM matrix elements leading to a SM predicted branching fraction of O($10^{-10}$).
The decay amplitude of $B^0_{s} \rightarrow \mu^{+} \mu^{-}$ can be enhanced significantly in some extensions
of the SM.
For instance, in the type-I\hspace{-0.1em}I two-Higgs-doublet-model (2HDM)~\cite{Log} the branching fraction is proportional
to ${\rm tan}^4\beta$, where ${\rm tan}\beta$ is the ratio between the vacuum expectation values of the two neutral Higgs fields.
In the minimal super-symmetric SM (MSSM)~\cite{Bab} the dependence on ${\rm tan}\beta$ is even stronger, $\propto$ ${\rm tan}^6\beta$.
Current observation of this decay at the Tevatron would necessarily imply new physics.

Current world best limit on  ${\mathcal B}(B_s^0\to\mu^+\mu^-)$ is given by CDF using 2~${\rm fb}^{-1}$ of Run~I\hspace{-0.1em}I data~\cite{btomumu}.
The sensitivity of the analysis was improved significantly from the previous result~\cite{PRL2} by increasing the
integrated luminosity of the event sample, using an enhanced muon selection, employing a
neural network (NN) classifier to separate signal from background, and performing the search in a
two dimensional grid in dimuon mass and NN space.
The observed event rates were consistent with SM background expectations.
Extracted 90\% (95\%) C.L. limit was
${\mathcal B}(B^0_s \to \mu^+ \mu^-) < 4.7~(5.8)\times 10^{-8}$.

\begin{figure}[htbp]
\center{
\psfig{figure=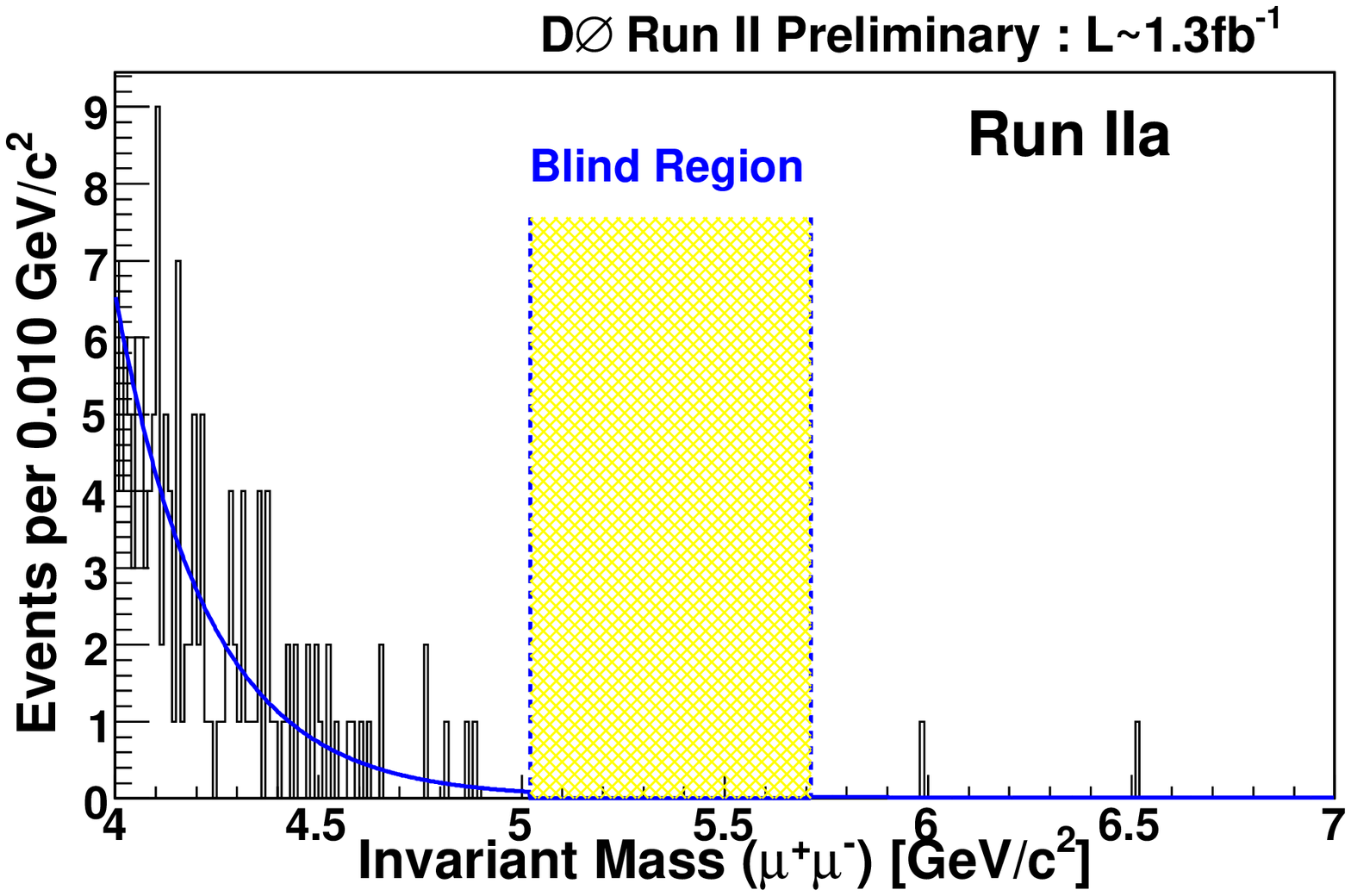,width=0.4\textwidth}
\psfig{figure=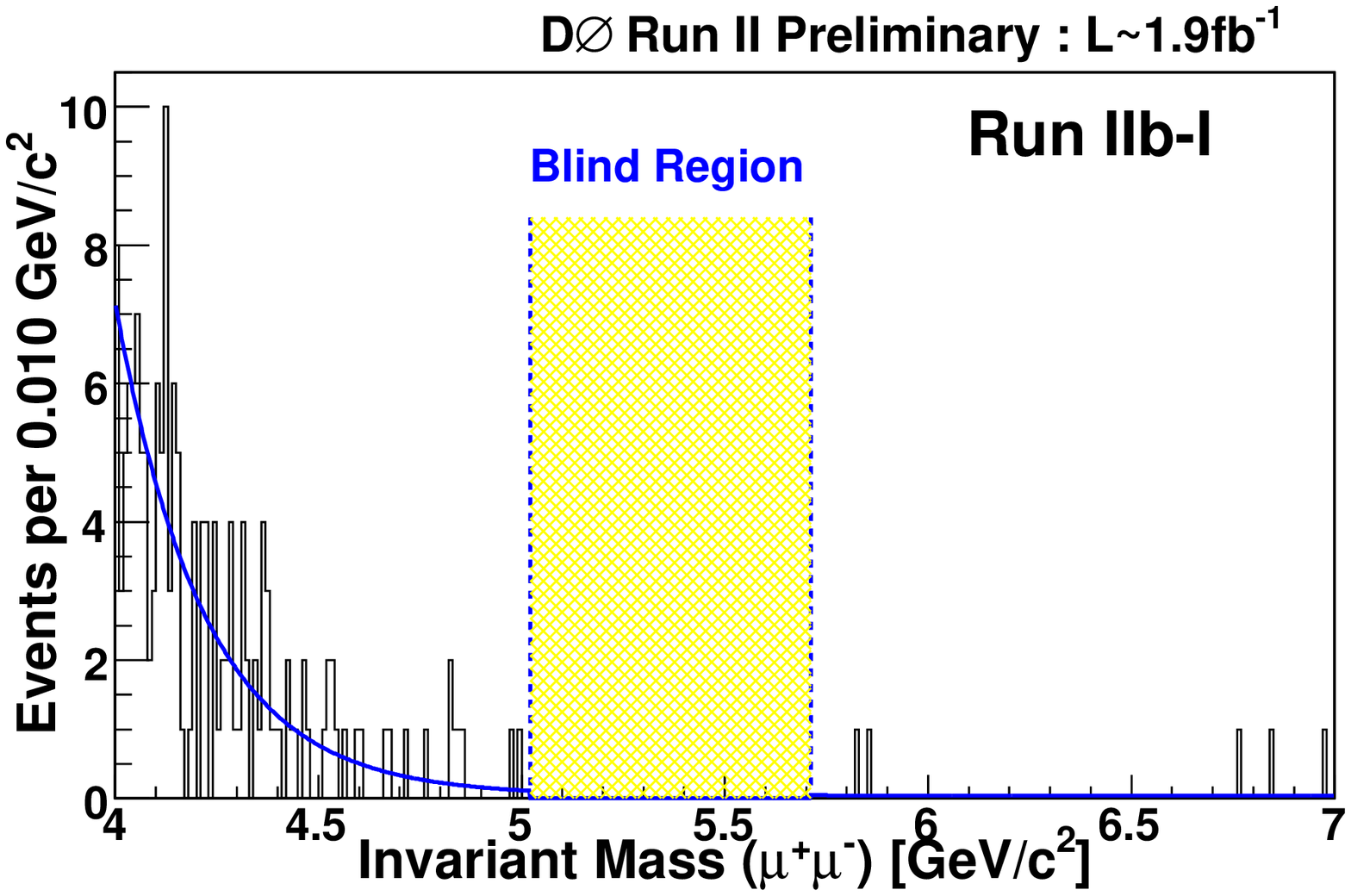,width=0.4\textwidth}
\begin{flushleft}
\hspace{1.4cm}
\psfig{figure=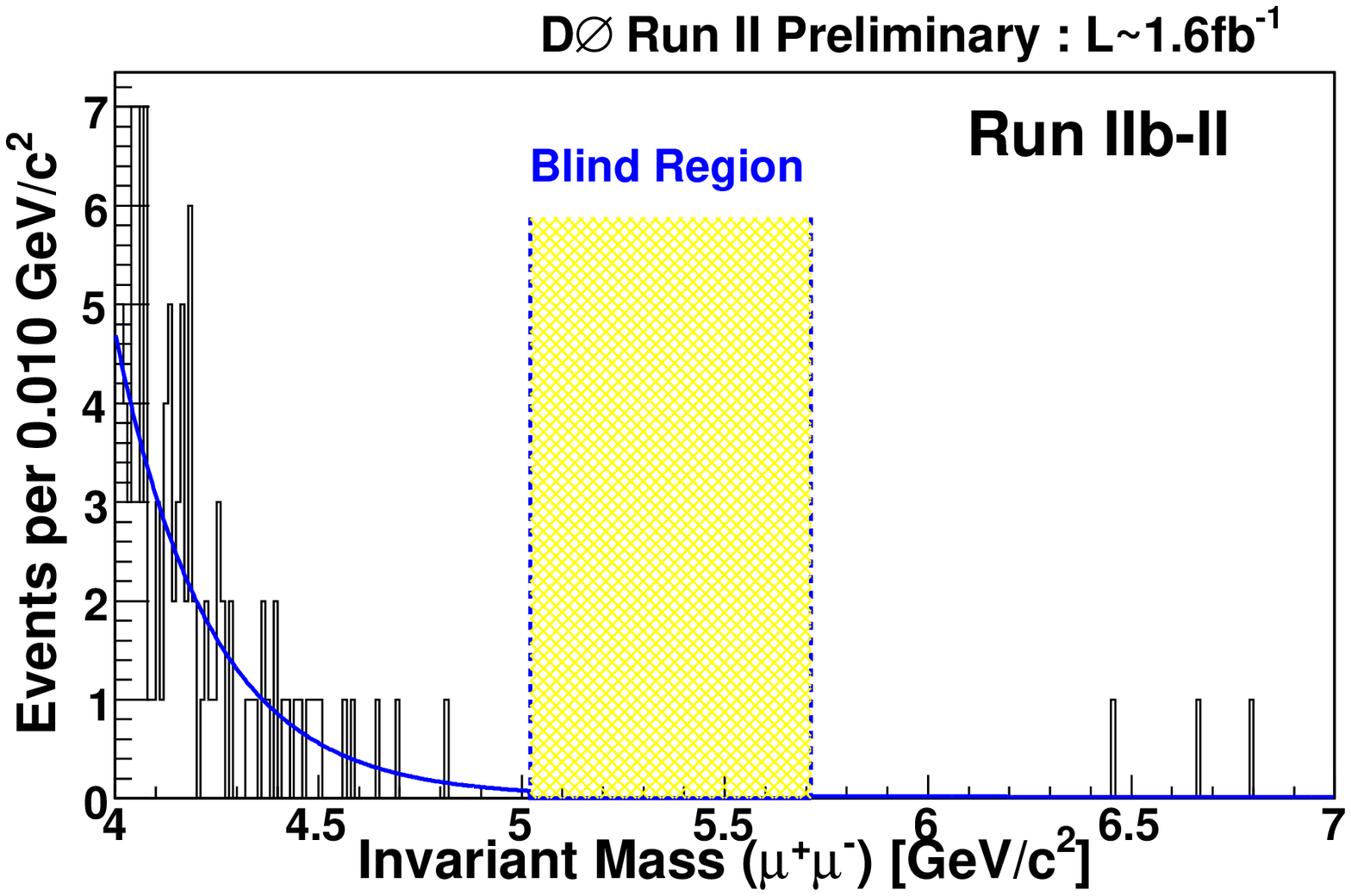,width=0.4\textwidth}
\end{flushleft}
\caption{Dimuon invariant mass distributions. Search box remains blinded.
\label{fig:bs2mm}}
}
\end{figure}

D\O\ reported a new expected upper limit on ${\mathcal B}(B^0_s \to \mu^+ \mu^-)$ using approximately $5~{\rm fb}^{-1}$ of dimuon trigger data~\cite{masato}.
The data sample was split into three subsamples, Run~I\hspace{-0.1em}Ia, Run~$\rm I\hspace{-0.1em}I$b-I and Run~I\hspace{-0.1em}Ib-I\hspace{-0.1em}I, based on the data taking period.
Roughly the integrated luminosity of Run~${\rm I\hspace{-0.1em}Ia}$ data is $1.3~{\rm fb}^{-1}$, Run~I\hspace{-0.1em}Ib-I is $1.9~{\rm fb}^{-1}$ and Run~I\hspace{-0.1em}Ib-I\hspace{-0.1em}I is $1.6~{\rm fb}^{-1}$.
The three subsamples were treated as three different and independent analyses, but the final upper limit was combined from the separate analyses.
To separate signal from background a boosted decision tree (BDT)~\cite{tmva} classifier was constructed, where five input variables of the $B^0_s$ meson
(isolation, transverse~momentum, transverse~decay~length~significance, impact~parameter~significance and logarithm~of~vertex~$\chi^2$~probability) were used.
Figure~\ref{fig:bs2mm} shows the dimuon invariant mass distributions after applying the BDT cut, where only events not in the blinded region $\pm 3 \sigma_m$ ($\sigma_m=0.115~{\rm GeV}/c^2$) around the $B^0_s$ mass are shown.
The same BDT cut was applied on the normalization channel, $B^+ \to J/\psi K^+$, and the following numbers of events were found: $1847 \pm 49(stat.) \pm 115(syst.)$ events in Run~$\rm I\hspace{-0.1em}I$a data, $2188 \pm 52(stat.) \pm 123(syst.)$ events in Run~I\hspace{-0.1em}Ib-I data
and $1683 \pm 46(stat.) \pm 112(syst.)$ events in Run~I\hspace{-0.1em}Ib-I\hspace{-0.1em}I data.
The systematic uncertainty is coming from our parametrization of the $B^{+}$ mass shape.
The random track contribution in the signal region $\pm2.5\sigma_m$ was estimated by extrapolating events in the sideband region to the signal region.
In addition,
possible non-negligible contributions of misidentified $B_s^0 \to K^+ K^-$ and $B^0 \to K^+ \pi^-$ were estimated.
The total background events in the search window in Run~I\hspace{-0.1em}Ia, Run~I\hspace{-0.1em}Ib-I, and Run~I\hspace{-0.1em}Ib-I\hspace{-0.1em}I data sets were estimated to be $2.16 \pm 0.62$, $3.73 \pm 1.07$ and $2.15 \pm 0.63$, respectively.
The expected SM yields of $B_s^0 \to \mu^+ \mu^-$ events in Run~I\hspace{-0.1em}Ia, Run~I\hspace{-0.1em}Ib-I, and Run~I\hspace{-0.1em}Ib-I\hspace{-0.1em}I data sets are $0.192 \pm 0.034$, $0.193 \pm 0.034$ and $0.139 \pm 0.025$, respectively.
Aside from the background uncertainty, the largest uncertainty common to the three data sets, 15.2\%, comes from the fragmentation ratio between $B^+$ and $B^0_s$.
Assuming no signal counts (background only) in the signal region, an expected upper limit on the branching fraction at the 90\%(95\%) C.L was computed.
The number of observed events was set to the number of background events, 2 events for Run~I\hspace{-0.1em}Ia, 4 events for Run~I\hspace{-0.1em}Ib-I and 2 events for Run~${\rm I\hspace{-0.1em}Ib}$-${\rm I\hspace{-0.1em}I}$.
In this calculation, it was assumed that there are no contributions from $B^0 \rightarrow \mu^+ \mu^-$ decays, where the decay is suppressed by $|V_{td}/V_{ts}|^2 \approx 0.04$.
The expected 90\%(95\%) upper limits for the branching fraction were $7.6~(9.4) \times 10^{-8}$ for Run~$\rm I\hspace{-0.1em}I$a data,
$9.9~(11) \times 10^{-8}$ for Run~$\rm I\hspace{-0.1em}I$b-I data and  $10~(13) \times 10^{-8}$ for Run I\hspace{-0.1em}Ib-I\hspace{-0.1em}I data.
The combined upper limit is then $4.3~(5.3) \times 10^{-8}$ at the 90\%(95\%) C.L.
This sensitivity is comparable with the best published upper limit from CDF~\cite{btomumu},
and improves the previous D\O\ result~\cite{d0_2fb} by a factor of two.
Further work to understand and reduce the background as well as to include more data is ongoing before opening the search box.

\section{Conclusion}

CDF has reported new upper limits on ${\mathcal B}(B^0_{(s)}\to e^+\mu^-)$ using $2~{\rm fb}^{-1}$ of data, and D\O\ has reported an new expected upper limit on ${\mathcal B}(B^0_s \to \mu^+\mu^-)$ using approximately $5~{\rm fb}^{-1}$ of data.
With the continuously increasing amount of data provided by the Tevatron and improvements of the analyses
measurements of rare $B$ decays provide new insight into the properties of the FCNC decays,
which allows for improved tests of the SM that could guide us to new physics scenarios.


\begin{thebibliography}{99}
\bibitem{IP} The impact parameter $d_0$ is the distance of closest approach of the track to the beam line.
\end{thebibliography}
\end{document}